\documentclass[12pt]{iopart}
\usepackage{graphicx}

\begin{document}
\title{Demonstration of the Universality of Molecular Structures in Prolate Deformed Nuclei}
\author{R. Canavan}
\address{Department of Physics,
	University of Surrey,
	Guildford,
	Surrey,
	GU2 7XH, U.K. }
\author{M. Freer}
\address{School of Physics and Astronomy, University of Birmingham, Edgbaston,
Birmingham~B15~2TT, U.K.}

\date{Received \today}

\begin{abstract}
The relationship between the deformed harmonic oscillator and the formation of molecular cluster structures, whereby valence neutrons are exchanged between cluster cores, is examined. It is found that there is a strong connection between the properties of the valence orbitals associated with deformed structures in the deformed harmonic oscillator and the molecular orbitals created by linear combinations of single-centre orbitals around nuclear clusters. The conclusion is that in addition to the appearance of clustering in the deformed harmonic oscillator that \emph{every} prolate deformed cluster structure should have molecular orbitals built on that structure. This is demonstrated through a series of examples that range from $^{13}$C to $^{57}$Ni. 
\end{abstract}
\maketitle

\section{Introduction}
The appearance of clusters in light nuclei is well documented~\cite{voe06,Fre18}. The most prominent include the $^{8}$Be ground-state and the $^{12}$C excited Hoyle-state. Both have $\alpha$-particle cluster structures and both play a key role in the synthesis of carbon in stars, through the triple-alpha process. There is good evidence for more complex cluster structures, where, for example, $^{20}$Ne may be described in in terms of $^{16}$O+$\alpha$ substructures and there is a long history of clustering in $^{24}$Mg, where $\alpha$+$^{16}$O+$\alpha$ clustering is one cluster configuration~\cite{voe06}. 

The formation of clusters presents an interesting possibility, which is well illustrated by the Beryllium isotopes. The $^8$Be nucleus is unbound to decay to two $\alpha$-particles in its ground-state, whereas $^9$Be is stable. The additional neutron has been found to exist in a delocalised orbit, covalently exchanged between the two $\alpha$-particle clusters. The existence of such structures was discussed by von Oertzen in a series of seminal papers~\cite{voe96,voe97}. These molecular orbitals in the Beryllium isotopes have both $\pi$ and $\sigma$ characteristics, where the notation follows from the atomic analogues. The experimental evidence for such structures comes from the associated rotational bands and decay properties of the states. Perhaps the best example is that of $^{10}$Be, which provides evidence for a molecular structure of two $\alpha$-particles and 2 $\sigma$ covalent neutrons~\cite{Fre06,Suz13}. 

The extension of these ideas to three centres and carbon isotopes was made by Milin and von Oertzen~\cite{Mil02}, where the states in $^{13}$C were analysed in terms of their potential molecular characteristics. Convincing evidence for the molecular structure of the carbon isotopes remains to be demonstrated, though there is significant contemporary interest in the molecular structure of $^{14}$C~\cite{Bib19}. Finally, the ideas of nuclear molecules were extended to asymmetric clusters by von Oertzen~\cite{voe01a} and Kimura~\cite{Kim07} through an analysis of the Neon isotopes.

These studies have provided a basis for the understanding of the link between the appearance of molecular orbitals and the underlying cluster structures. This was further cemented through the study of the relationship between the states of the deformed harmonic oscillator and molecular states created through the H\"{u}ckle approach~\cite{McE04}. These molecular structures are found not only in such relatively schematic approaches, but also in more sophisticated methods that treat the nucleons individually, do not impose the clustering a priori and deal with realistic nucleon-nucleon interactions. There are many such approaches, see Ref.~\cite{Fre18} for a review, but the Antisymmeterised Molecular Dynamics, AMD, method has been particularly successful in both reproducing the molecular structures and experimental data, providing a level of confidence that the molecular degrees of freedom are a good representation of the structure of a number of light nuclei.

The present work builds on Ref.~\cite{McE04}, which examined the molecular states of $\alpha$-like nuclei, and extends this understanding to the full spectrum of prolate deformed cluster states across the entire mass range of possible cluster structures.

\section{Theoretical Approach}
The aim of the present paper is to explore the nature of molecular wave-functions across a wide range of nuclear systems. The framework for this is the deformed harmonic oscillator, DHO. The DHO solutions for molecular-like nuclei are  compared with results of calculations where the molecular structures are created explicitly from the orbits of linear combinations of two-centre orbits created using the H\"{u}ckle method. 

\begin{figure}
  \begin{center}
  \includegraphics[width=0.38\textwidth]{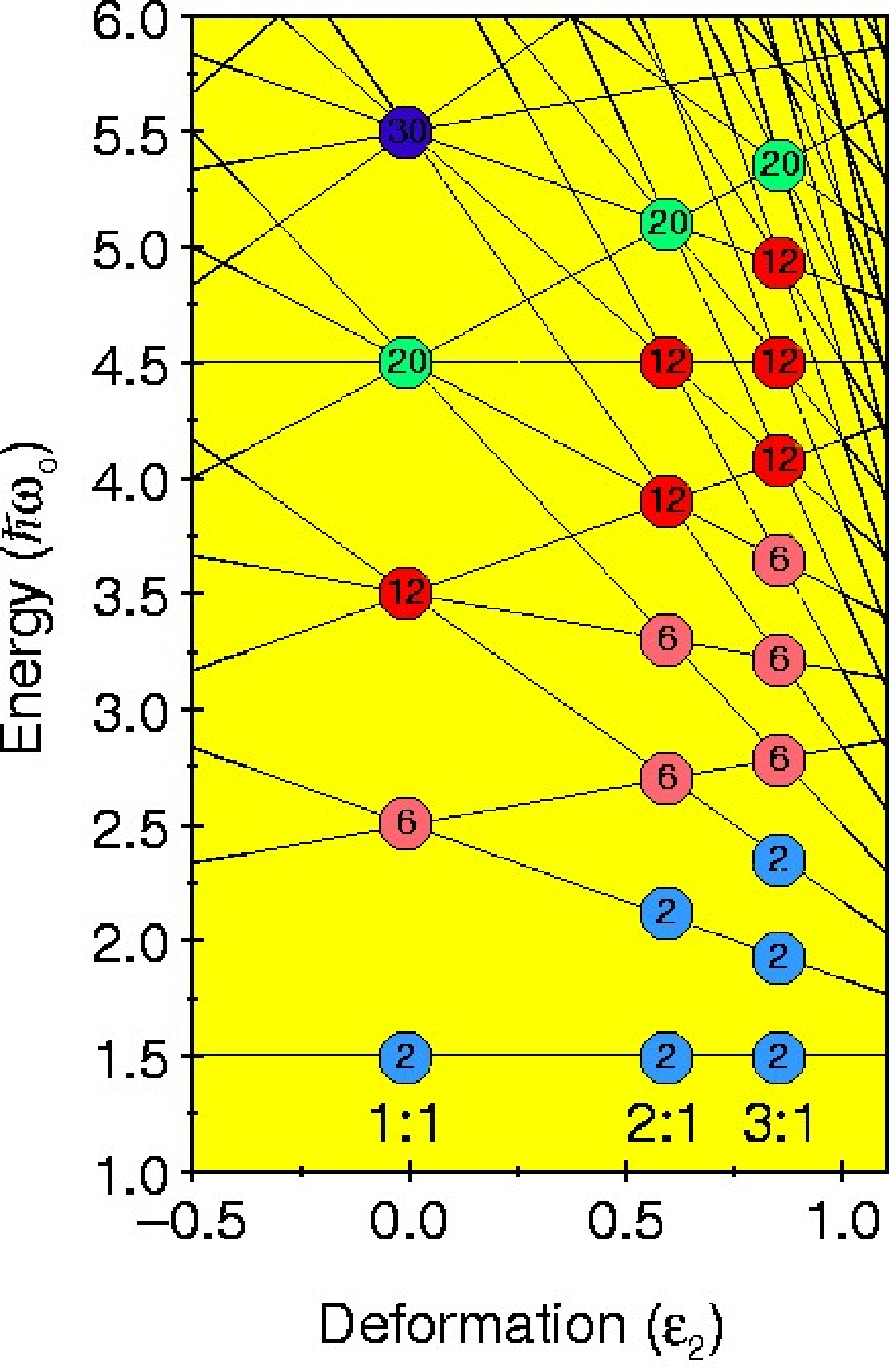}
    \vspace{-0.0cm}
      \hspace{-0.3cm}
      \caption{Energy levels of the deformed axially symmetric
                 harmonic oscillator as a function
                 of the  quadrupole deformation (oblate and prolate, i.e. negative
                 and positive values of $\epsilon_2$, respectively).
                 Degeneracies (labelled) appear due to crossings of orbits at
                 certain ratios of the length of the long axis (the symmetry axis)
                 to the short perpendicular  axis.
                The regions of high degeneracy define a  shell closure
                 for deformed shapes \cite{Fre95}.}
    \label{fig:HO}
  \end{center}
\end{figure}

\subsection{Deformed Harmonic Oscillator}
The DHO provides a good description of light nuclear systems, where the inclusion of the spin-orbit interaction and an $l^2$ dependence of the nuclear potential results in the Nilsson description of deformed nuclei~\cite{Fre95}. The deformed harmonic oscillator has the well-known shell-like structure where at integer ratios of axial deformations the shell structure is enhanced,~Fig.~\ref{fig:HO}. This shell structure may be associated with stable deformed structures and indeed cluster partitions~\cite{Rae89}.

 The total energy at each level for the DHO is given by:
\begin{equation}
E = \hbar \omega_{x} n_{x} + \hbar \omega_{y} n_{y} + \hbar \omega_{z} n_{z} + \frac{3}{2} \ \hbar \omega_{0},
\end{equation}
where $\omega_{x}$, $\omega_{y}$ and $\omega_{z}$ and $n_{x}$, $n_{y}$ and $n_{z}$ are the oscillation frequencies and number of oscillation quanta in the $x$, $y$, and $z$ directions respectively, and $\omega_{0}$ is the average oscillation frequency. Furthermore, the oscillation frequencies must satisfy:
\begin{equation}
\omega_{0}^3 = \omega_{x}  \omega_{y}  \omega_{z}.
\end{equation}
Therefore, for a nucleus with, for example, a 2:1 deformation along the $z$-direction, the energy levels are given by:
\begin{equation}
E = \frac{6}{5} \ \hbar \omega_{0} n_{x} + \frac{6}{5} \ \hbar \omega_{0} n_{y} + \frac{3}{5} \ \hbar \omega_{0} n_{z} + \frac{3}{2} \  \hbar \omega_{0}.
\end{equation}
The orbitals in the DHO are labelled by [$n_{x}$, $n_{y}$, $n_{z}$], and the number of oscillator quanta in each Cartesian direction determines the orientation of that orbital.

The association of cluster configurations with the shell structure~\cite{Fre95,Rae89} follows from the degeneracies illustrated in Fig.~\ref{fig:HO}. Within the DHO the nucleus $^8$Be, for example, is created through the population of the DHO orbitals,  [$n_{x}$, $n_{y}$, $n_{z}$]$=$[0,0,0]$^4$ and [0,0,1]$^4$.
At a 2:1 deformation this is seen to be associated with the filling of the two orbitals with 2$p$+2$n$, i.e. 2$\alpha$-particles, and hence $^8$Be may be thought of as 2 $\alpha$-clusters. The densities calculated from this DHO configuration also reveal the two-centre cluster structure~\cite{Fre95}. For the next most complex case which is $^{20}$Ne, this also corresponds to a 2:1 shell gap. Here the degeneracies are 2+(2+6) and may be linked to a $^4$He+$^{16}$O cluster structure. The next shell gap at a deformation of 2:1 occurs for degeneracies 2+2+8+8 which would correspond to a 2:1 deformation of $^{32}$S and a $^{16}$O+$^{16}$O cluster structure. A similar situation occurs at a deformation of 3:1, for example the shell gap at cumulative degeneracies of 2+2+2+6+6+6 being linked to the 3:1 deformed state in $^{48}$Cr and a $^{16}$O+$^{16}$O+$^{16}$O cluster configuration.

To form molecular structures in these cluster-systems where the valence particle, typically a neutron, is exchanged between the cluster cores, one needs to examine the properties of the next available orbits beyond those associated with the $\alpha$ and $^{16}$O cluster cores. These are then those which exist immediately above the shell gap. In the case of $^8$Be+$n$ then these are the three levels [1,0,0], [0,1,0] and [0,0,2]. For $^{32}$S it is the 6 orbitals [1,1,0], [2,0,0], [0,2,0], [1,0,2], [0,1,2] and [0,0,4]. In order to ascertain if these have a molecular character, their properties need to be compared with those created explicitly from the available two-centre orbitals. The present work has been framed in terms of neutron covalent orbitals as these are the most prevalent in terms of experimental counterparts. However, equally the ideas may be applied to valence protons or combinations of protons and neutrons.

\subsection{Neutron Wave-functions in Two-Centred Systems}

\paragraph{}The neutron wave-functions may be created from linear combinations of the single-centre obits of the valence neutron around the cluster core. In the case of the $\alpha$-particle the valence orbitals are, [$n_{x}$, $n_{y}$, $n_{z}$]=[1,0,0], [0,1,0] and [0,0,1]. For $^{16}$O, the valence orbitals are the degenerate (in the spherically symmetric HO) [2,0,0], [0,2,0], [0,0,2], [1,1,0], [1,0,1] and [0,1,1] levels. Wave-functions can be linearly combined to create the multiple centre case to compare with the orbital wave-functions found in the DHO. Hence, the nuclear molecular orbit ($\psi$) is then given by a sum of the neutron wave-functions ($\phi$) at each cluster centre, with the appropriate normalisation constants, $c_i$, (determined by the number of cluster centres in the molecule):
\begin{equation}
\psi = c_{1} \phi_{1} + c_{2} \phi_{2} + c_{3} \phi_{3} + \dots
\end{equation}

To find the relative contribution which each single-centre wave-function makes to the overall molecular one,  the H\"{u}ckel method was followed. The appropriate coefficients are found by solving the secular determinant equation which arises from the Schr\"{o}dinger equation for the nuclear molecule (Section~\ref{sec:Huc}).

When making linear combinations of the orbitals in the multiple centre model, the quantum numbers defining the orbital obey certain rules which allowed the combination to be associated with one of the single-centre orbitals~\cite{Fre07, McE04}:
\begin{itemize}
 \item The $n_z$ quantum numbers of the single centre orbits should be summed over the number of centres in the molecule,  $C$, i.e. $n_z=\sum^C_i n_{zi}$, to form the lowest energy molecular configuration.
  \item The $n_x$ and $n_y$ quantum numbers, perpendicular to the direction of deformation, are conserved; the value for $n_{x}$ and $n_{y}$ remains unchanged in the composite system for symmetric clusters ($n_{x1}=n_{x2}=n_{x}$ and $n_{y1}=n_{y2}=n_{y}$).
  \item When forming the orbitals for the  excited molecular states, the quantum number along the direction of deformation can be found using:
  $n_{z} = \sum^{C}_{i} n_{zi} + C - 1$
  where $n_{z}$ is the quantum number for the orbital in the molecule, $n_{zi}$ are the quantum numbers at each centre in the multiple centre molecule.
\end{itemize}

\subsection{The H\"{u}ckel Method}
\label{sec:Huc}
The wave-functions describing all quantum particles can be found by solving the Schr\"{o}dinger equation, for the molecular system
\begin{equation}
\hat{H} \psi = E \psi
\end{equation}
where $\hat{H}$ is the Hamiltonian of the nucleus, $E$ is the energy eigenvalue associated with the particular molecular state, and $\psi$ is the overall wave-function of the nuclear molecule as given above. To demonstrate the method the case of $^{33}$S will be considered. Here there are two $^{16}$O clusters and a single valence neutron. The wave-function describing the valence neutron $^{33}$S is then:
\begin{equation}
\psi = c_{1} \phi_{1} + c_{2} \phi_{2}
\label{eq:wavefunc}
\end{equation}
$\phi_{1}$ and $\phi_{2}$ are the wave-functions of the neutron orbitals at each cluster centre (given by the single centre [2,0,0], [0,2,0], [0,0,2], [1,1,0], [1,0,1] and [0,1,1] orbitals), and the coefficients $c_{1}$ and $c_{2}$ give their relative contributions to the overall wave-function of the nucleus. Equation \ref{eq:wavefunc} is then multiplied by the conjugate wave-function $\psi^{*}$ and integrated over all space, a rearrangement of the equation then gives:
\begin{equation}
E = \frac{\int \psi^{*} \hat{H} \psi \ d\tau}{\int \psi^{*} \psi \ d\tau}
\end{equation}
Substituting the molecular wave-function of $^{33}$S into its eigenvalue equation yields:
\begin{equation}
E = \frac{c^{2}_{1} H_{11} + 2 \ c_{1}c_{2}H_{12} + c^{2}_{2}H_{22}}{c^{2}_1 + 2 \ c_{1}c_{2}S_{12} + c^{2}_{2}}
\end{equation}
Where the equation has been simplified using $H_{ij} = \int \psi_{i}^{*} \hat{H} \psi_{j} \ d\tau$ and $S_{ij} = \int \psi_{i}^{*} \psi_{j} \ d\tau$. The actual wave-function of the nuclear molecule, given by the values of $c_{1}$ and $c_{2}$ will yield the lowest energy state (rather than an excited state); therefore the coefficients can be found by minimising the energy eigenvalue with respect to $c_{1}$ and $c_{2}$. The following secular equations are obtained by finding the partial derivatives and setting them equal to zero:
\begin{equation}
c_{1} \ (H_{11} - E) + c_{2} \ (H_{12} - E S_{12}) = 0,
\end{equation}
\begin{equation}
c_{1} \ (H_{12} - E S_{12}) + c_{2} \ (H_{22} - E) = 0.
\end{equation}
These can be equivalently written in matrix form:
\begin{equation}
\left(
\begin{array}{c c}
H_{11} - E & H_{12} - E S_{12} \\
H_{12} - E S_{12} & H_{22} - E
\end{array}
\right)
\left(
\begin{array}{c}
c_{1} \\ c_{2}
\end{array}
\right) = \left(\begin{array}{c}
0 \\ 0
\end{array}
\right).
\end{equation}
The energy eigenvalue, $E$, which satisfies this equation is found by solving the secular determinant, because for a non-trivial solution the matrix must be singular:
\begin{equation}
det \left(
\begin{array}{c c}
H_{11} - E & H_{12} - E S_{12} \\
H_{12} - E S_{12} & H_{22} - E
\end{array}
\right) = 0.
\end{equation}

The secular determinant equation may be simplified by making assumptions about the interactions within the nuclear molecule. $H_{11}$ and $H_{22}$ represent the self-interaction energies of the free $^{17}$O nuclei at each centre (i.e. the $n$+$^{16}$O binding energy), which are identical due to the symmetry of the molecule, this value is set equal to $\alpha$. The interaction energy between the two clusters is given by $H_{12} = H_{21} = \beta$ if the $^{16}$O clusters are neighbouring (as is the case for $^{32}$S) or set to zero otherwise. Additionally, due to the neutron wave-functions $\phi_{1}$ and $\phi_{2}$ being located at different cluster centres, their separation enables the simplifying assumption of small overlap to be made, so $S_{12} = S_{21} \simeq 0$. Finally, if the neutron wave-functions at each cluster centre have been normalised then $S_{11} = S_{22} = 1$. Using all of these assumptions gives the secular determinant equation:
\begin{equation}
det \left(
\begin{array}{c c}
\alpha - E & \beta \\
\beta & \alpha - E
\end{array}
\right) = 0.
\end{equation}

The solution to this equation gives $E = \alpha \pm \beta$, and
\begin{equation}
c_{1} \ (\alpha - E) + c_{2} \ \beta = 0 \\
c_{1} \ \beta + c_{2} \ (\alpha - E) = 0,
\end{equation}
the solutions for which are $c_{1}$ = $c_{2}$ and $c_{1}$ = $-\ c_{2}$.

As the molecular wave-function must be normalised, for $^{33}$S the solutions are:
\begin{equation}
\psi = \frac{1}{\sqrt{2}} \ \Big(\phi_{1} \pm \phi_{2}\Big).
\end{equation}
The same method for calculating the molecular wave-function for nuclei comprising more complex arrangements of clusters, i.e. for 3 or more centre systems. The wave-function for each neutron orbital can be constructed in the same way (by solving the secular determinant equation), and linear combinations were made by using the appropriate coefficients. The coefficients for collinear molecules (i.e. when the clusters form a 1d linear structure) of up to six clusters are given in Table \ref{tab1}~\cite{McE04}.

For the two-centre system molecular wave-function is given by:
\begin{equation}
\psi(r) = \frac{1}{\sqrt{2}} \ \Bigg( \frac{1}{\sqrt{\pi^{1/2} 2^{n} n!}}H_{n}(r_{1})exp\left(\frac{- r_{1}^{2}}{2}\right) \pm \frac{1}{\sqrt{\pi^{1/2} 2^{n} n!}}H_{n}(r_{2})exp\left(\frac{- r_{2}^{2}}{2}\right) \Bigg)
\end{equation}
Here, the HO scale factor $\sqrt{m\omega/\hbar}$ is set to unity and $n$ denotes the number of oscillator quanta, $r_{1}$ and $r_{2}$ give the locations of the two clusters with respect to the origin and $H_{n}$ is the appropriate Hermite polynomial. For this investigation, the wave-functions were projected onto the $y-z$ plane and integrated up over the $x$ dimension, and the $z$-direction was chosen to be the deformation axis, or the axis of separation of the clusters.

\begin{table} 
\caption{The coefficients of the single particle wave-functions in
the molecular orbits of {\it prolate} nuclei. The magnitude of the
coefficients determine the relative contribution of the
wave-function to the particular molecular orbit, taken from~\cite{McE04}. The increasing orbit number corresponds to higher number of nodes in the composite wave-function and hence higher energy levels.}\label{tab1}
\begin{center}
\begin{tabular} {ccccccc} \hline
 Orbit ($i$) & C$_{i}$ & $^{9}$Be   & $^{13}$C & $^{17}$O & $^{21}$Ne & $^{25}$Mg \\
 \hline
1 & 1 & 0.707           & 0.5       & 0.372   & 0.289 & 0.232    \\
  & 2 & 0.707           & 0.707     & 0.602   & 0.5   & 0.418    \\
  & 3 &                 & 0.5       & 0.602   & 0.577 & 0.521    \\
  & 4 &                 &          & 0.372    & 0.5   & 0.521    \\
  & 5 &                 &          &          & 0.289 & 0.418    \\
  & 6 &                 &          &          &          & 0.232    \\
\hline
2 & 1 & 0.707           & 0.707    & $-$0.602 & 0.5      & $-$0.418 \\
  & 2 & $-$0.707        & 0        & $-$0.372 & 0.5      & $-$0.521 \\
  & 3 &                 & $-$0.707 & 0.372    & 0        & $-$0.232 \\
  & 4 &                 &          & 0.602    & $-$0.5   & 0.232    \\
  & 5 &                 &          &          & $-$0.5   & 0.521    \\
  & 6 &                 &          &          &          & 0.418    \\
\hline
3 & 1 &                 & 0.5   & 0.602    & $-$0.577 & 0.521    \\
  & 2 &                 & $-$0.707    & $-$0.372 & 0        & 0.232    \\
  & 3 &                 & 0.5   & $-$0.372 & 0.577    & $-$0.418 \\
  & 4 &                 &          & 0.602    & 0        & $-$0.418 \\
  & 5 &                 &          &          & $-$0.577 & 0.232    \\
  & 6 &                 &          &          &          & 0.521    \\
\hline
4 & 1 &                 &          & $-$0.372 & 0.5      & $-$0.521 \\
  & 2 &                 &          & 0.602    & $-$0.5   & 0.232    \\
  & 3 &                 &          & $-$0.602 & 0        & 0.418    \\
  & 4 &                 &          & 0.372    & 0.5      & $-$0.418 \\
  & 5 &                 &          &          & $-$0.5   & $-$0.232 \\
  & 6 &                 &          &          &          & 0.521    \\
\hline
5 & 1 &                 &          &          & $-$0.289 & 0.418    \\
  & 2 &                 &          &          & 0.5      & $-$0.521 \\
  & 3 &                 &          &          & $-$0.577 & 0.232    \\
  & 4 &                 &          &          & 0.5      & 0.232    \\
  & 5 &                 &          &          & $-$0.289 & $-$0.521 \\
  & 6 &                 &          &          &          & 0.418    \\
\hline
6 & 1 &                 &          &          &          & $-$0.232 \\
  & 2 &                 &          &          &          & 0.418    \\
  & 3 &                 &          &          &          & $-$0.521 \\
  & 4 &                 &          &          &          & 0.521    \\
  & 5 &                 &          &          &          & $-$0.418 \\
  & 6 &                 &          &          &          &
  0.232 \\
  \hline
\end{tabular}
\end{center}
\end{table}
\section{Results}

\begin{figure}
  \begin{center}
  	\vspace{-4cm}
  \includegraphics[width=0.8\textwidth]{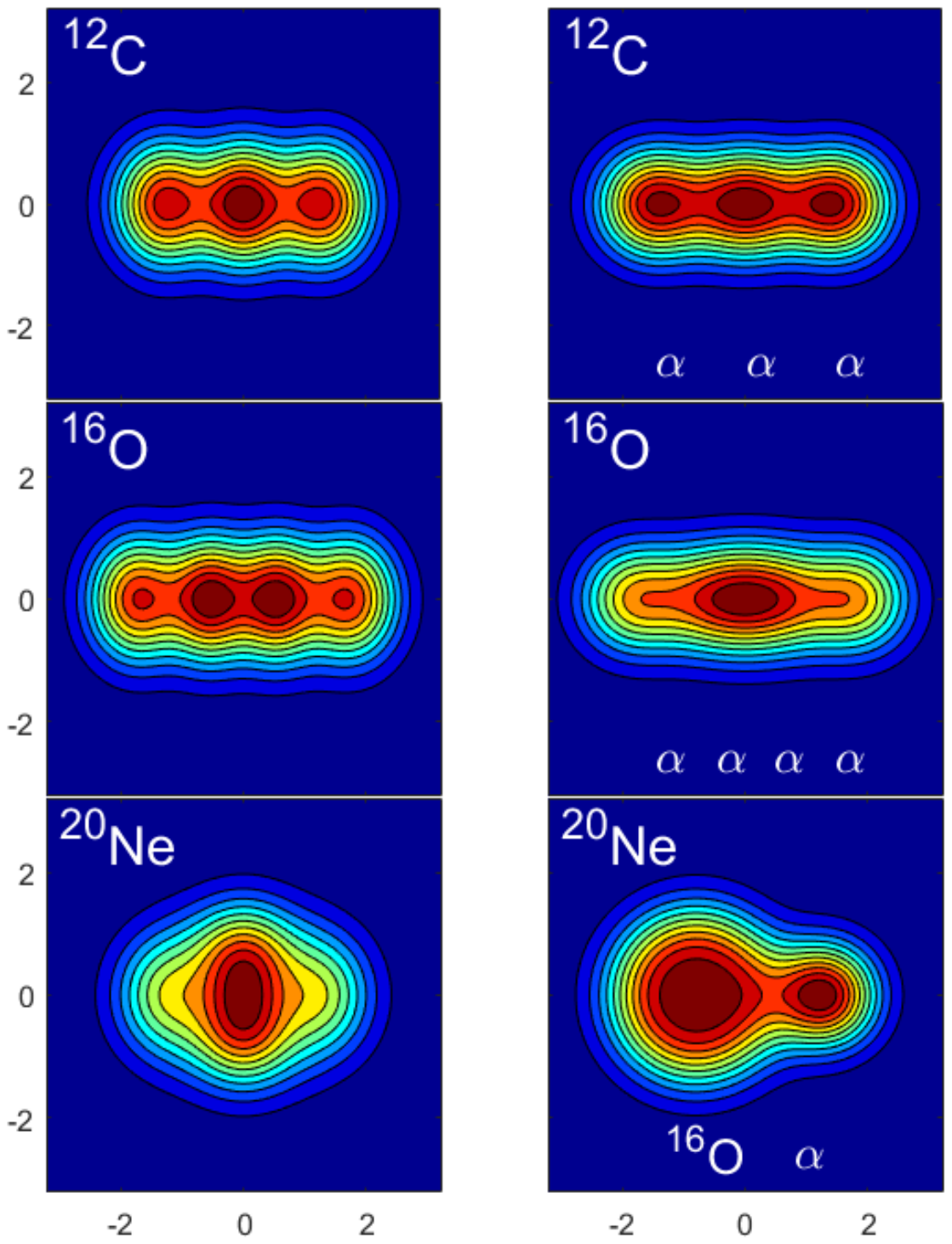}
    \vspace{-4cm}
      \hspace{-0.3cm}
      \caption{\emph{(left)} The deformed harmonic oscillator density calculations for the three nuclei $^{12}$C, $^{16}$O and $^{20}$Ne for the deformations 3:1, 4:1 and 2:1, respectively. \emph{(right)} The calculated densities for the three nuclei $^{12}$C, $^{16}$O and $^{20}$Ne formed from superimposing the densities of the individual clusters, i.e. 3$\alpha$, 4$\alpha$ and $^{16}$O+$\alpha$, respectively.    } 
    \label{fig:dho1}
  \end{center}
\end{figure}
\begin{figure}
  \begin{center}
  	\vspace{-4cm}
  \includegraphics[width=0.8\textwidth]{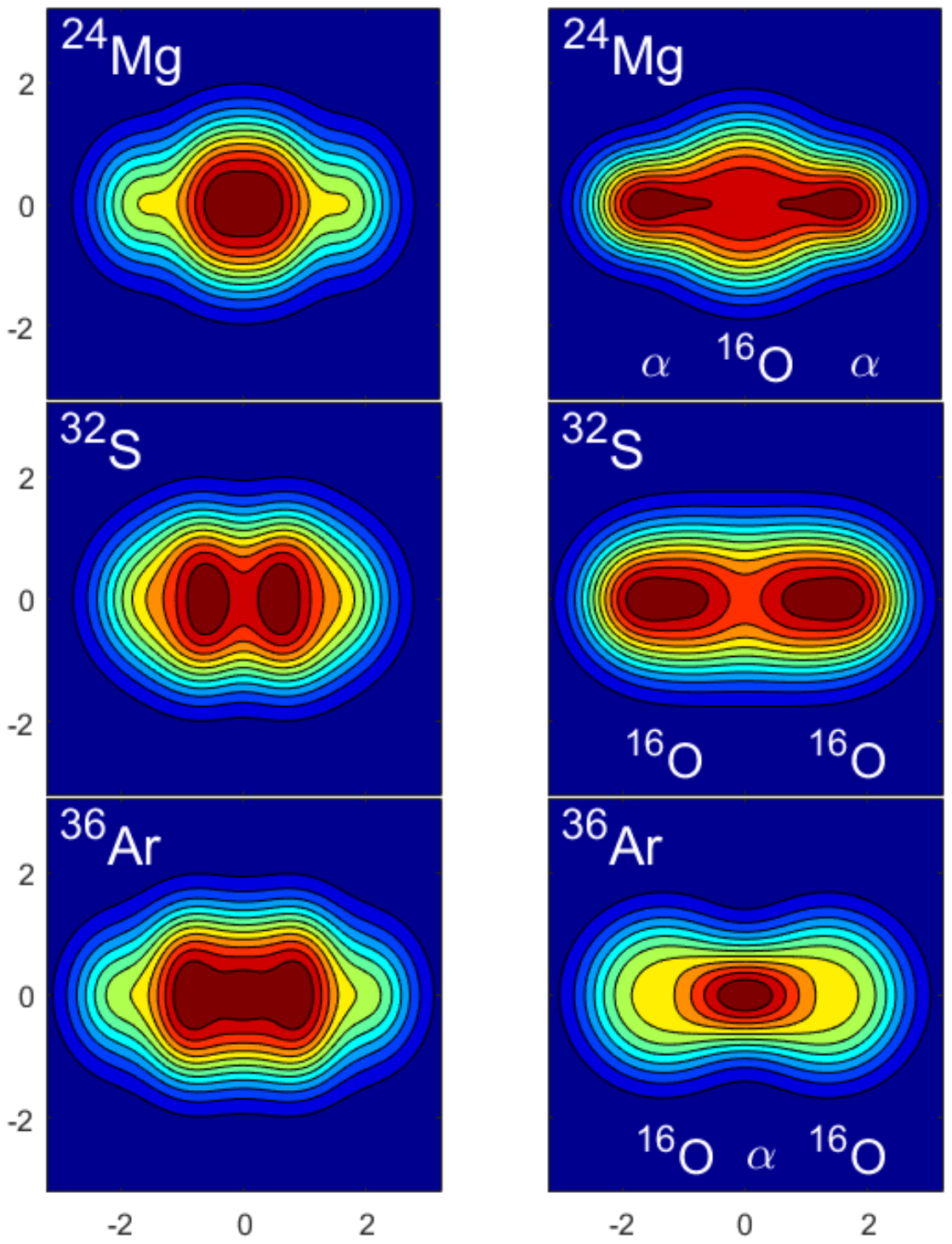}
    \vspace{-4cm}
      \hspace{-0.3cm}
      \caption{\emph{(left)} The deformed harmonic oscillator densities for the nuclei $^{24}$Mg, $^{32}$S and $^{36}$Ar, which reveal the underlying cluster structures $\alpha$+$^{16}$O+$\alpha$, $^{16}$O+$^{16}$O and $^{16}$O+$\alpha$+$^{16}$O, respectively. \emph{(right)} The densities of the three cluster structures $\alpha$+$^{16}$O+$\alpha$, $^{16}$O+$^{16}$O and $^{16}$O+$\alpha$+$^{16}$O created by superimposing the corresponding densities.    }
    \label{fig:dho2}
  \end{center}
\end{figure}
\begin{figure}
  \begin{center}
  	\vspace{-4cm}
  \includegraphics[width=0.8\textwidth]{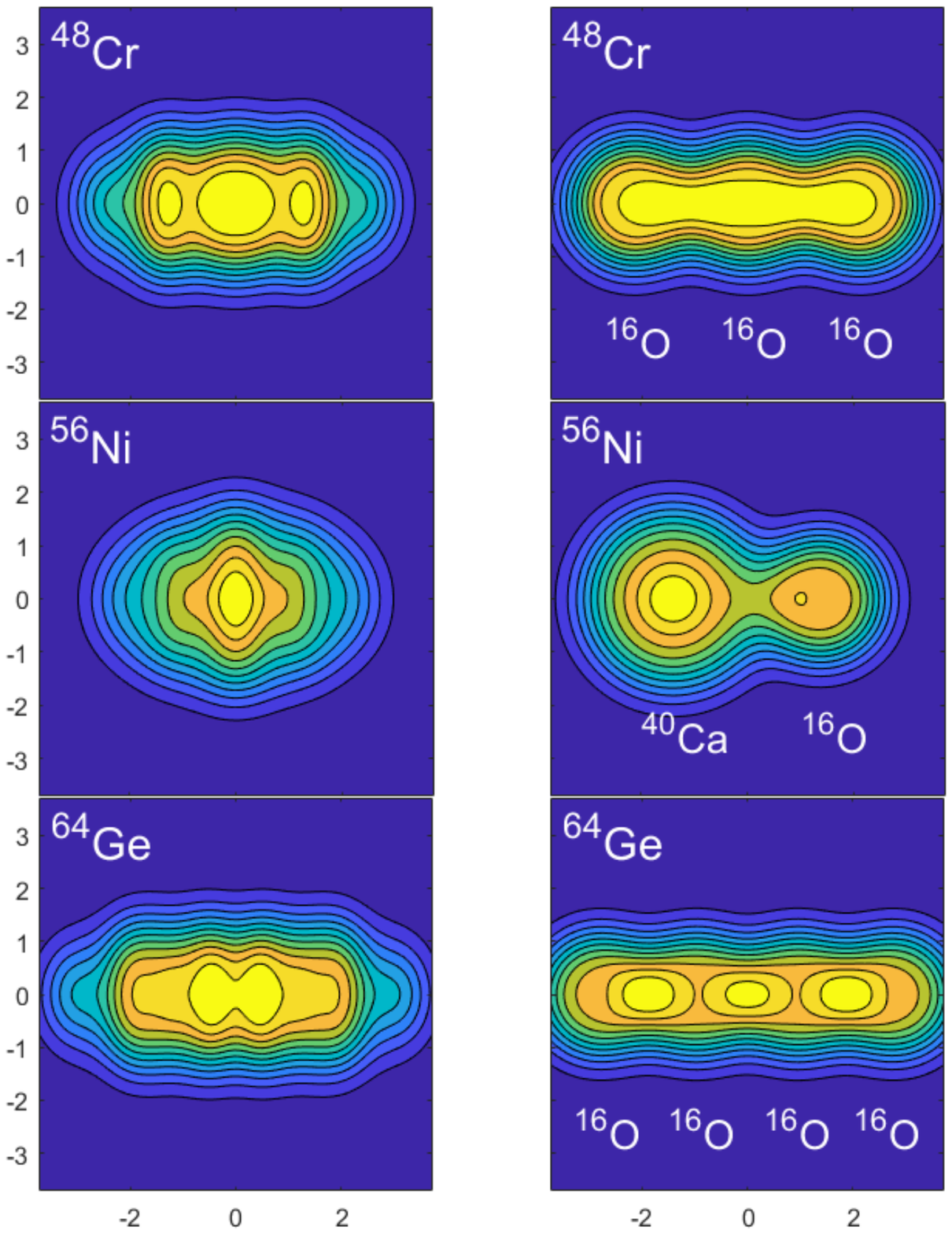}
   \vspace{-4cm}
      \hspace{-0.3cm}
      \caption{\emph{(left)} The deformed harmonic oscillator densities for the nuclei $^{48}$Cr, $^{56}$Ni and $^{64}$Ge, which reveal the underlying cluster structures $^{16}$O+$^{16}$O+$^{16}$O, $^{40}$Ca+$^{16}$O (asymmetric) and $^{16}$O+$^{16}$O+$^{16}$O+$^{16}$O, respectively. \emph{(right)} The densities of the three cluster structures $^{16}$O+$^{16}$O+$^{16}$O, $^{40}$Ca+$^{16}$O (asymmetric) and $^{16}$O+$^{16}$O+$^{16}$O+$^{16}$O created by superimposing the corresponding densities.}
    \label{fig:dho3}
  \end{center}
\end{figure}
\begin{figure}
  \begin{center}
  	\vspace{-4cm}
  \includegraphics[width=0.8\textwidth]{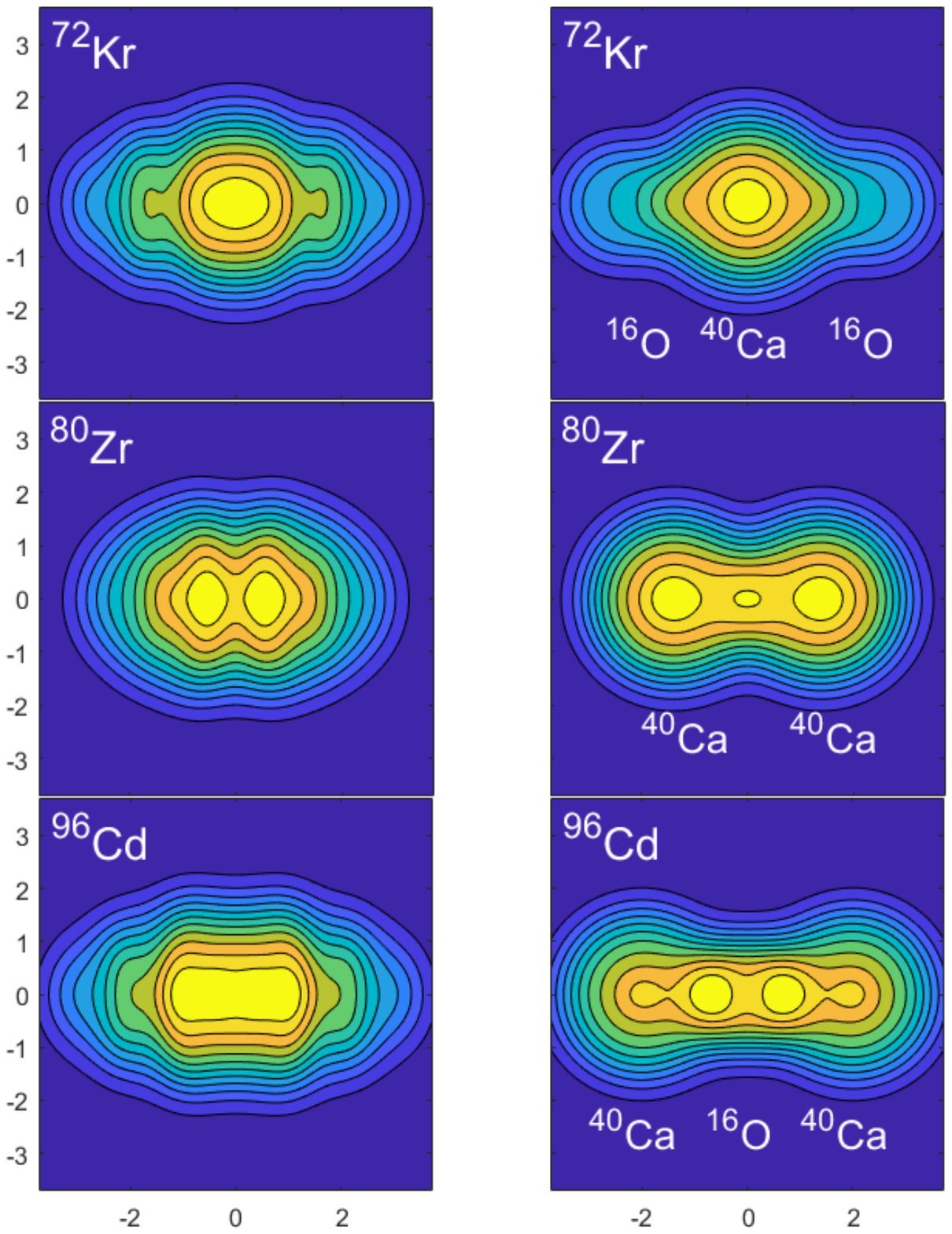}
   \vspace{-4cm}
      \hspace{-0.3cm}
      \caption{\emph{(left)} The deformed harmonic oscillator densities for the nuclei $^{72}$Kr, $^{80}$Zr and $^{96}$Cd, which reveal the underlying cluster structures $^{16}$O+$^{40}$Ca+$^{16}$O, $^{40}$Ca+$^{40}$Ca and $^{40}$Ca+$^{16}$O+$^{40}$Ca, respectively. \emph{(right)} The densities of the three cluster structures $^{16}$O+$^{40}$Ca+$^{16}$O, $^{40}$Ca+$^{40}$Ca and $^{40}$Ca+$^{16}$O+$^{40}$Ca created by superimposing the corresponding densities.}
    \label{fig:dho4}
  \end{center}
\end{figure}
Figures~\ref{fig:dho1} to ~\ref{fig:dho4} show the densities of the cluster cores built from both the DHO (single-centred nuclear core) and also by explicitly creating a superposition of the densities from the cluster components in a multi-centre approach. These cores correspond to the underlying structures around which the molecular neutrons would orbit. What is clear is that the symmetries that are found in the case of the degeneracies of the DHO (Fig.~\ref{fig:HO}) are broadly reproduced in terms of the densities, revealed in the multi-centre calculations. The multi-centre calculations were performed for separations of the cluster components which are approximately equivalent to the peaks in the density distributions indicated in the DHO densities. The oscillator scale parameter was adjusted from unity for the individual clusters according to the relationship $m\omega/\hbar=0.9A^{1/3}+0.7$ fm$^2$. It can be observed that where the densities of the clusters in the multicentre calculations overlap the densities increase and the details of the multicentre structure are diminished. This can be seen, for example, in the case of the 4$\alpha$ structure of $^{16}$O, in Fig.~\ref{fig:dho1}, where the 4$\alpha$ clusters are most evident in the low density region. However, in the deformed harmonic oscillator the cluster structure is often more evident, demonstrating the importance of the Pauli Exclusion Principle, PEP, in enhancing the cluster structure. Here the overlapping densities from the separated cluster cores does not obey the PEP. This is explored further in the latter part of this paper.

Fig.~\ref{fig:dho1} shows the two sets of calculations for the three isotopes $^{12}$C, $^{16}$O and $^{20}$Ne, which demonstrate the expected cluster structure associated with the 3:1, 4:1 and 2:1 shell closures, respectively. In the case of $^{20}$Ne the multi-centred calculations are for $^{16}$O+$\alpha$  which is reflection antisymmetric.  Similarly Fig.~\ref{fig:dho2} shows the $\alpha$-$^{16}$O-$\alpha$, $^{16}$O+$^{16}$O and $^{16}$O+$\alpha$+$^{16}$O structures in $^{24}$Mg, $^{32}$S and $^{36}$Ar, respectively, associated with the 3:1, 2:1 and 3:1 shell gaps. Evidence for the cluster structure in $^{48}$Cr ($^{16}$O+$^{16}$O+$^{16}$O), $^{56}$Ni ($^{40}$Ca+$^{16}$O), $^{64}$Ge ($^{16}$O+$^{16}$O+$^{16}$O+$^{16}$O), $^{72}$Kr ($^{16}$O+$^{40}$Ca+$^{16}$O), $^{80}$Zr ($^{40}$Ca+$^{40}$Ca) and $^{96}$Cd ($^{40}$Ca+$^{16}$O+$^{40}$Ca) is also found, Fig.~\ref{fig:dho4}, demonstrating that the DHO produces the cluster symmetries. There is no significant reduction of the cluster structure in the deformed harmonic oscillator even in the heaviest systems. As noted, it is these clustered structures which then create the cores for the valence neutrons, covalently exchanged between the clusters.

\subsection{Symmetric Two Centre Molecules}
The $^{9}$Be nucleus has been shown previously to contain two alpha clusters plus a covalent neutron~\cite{McE04}. The molecular characteristics are also found in the AMD calculations of the Beryllium isotopes, described at length in Ref.~\cite{Fre18}. The next two-centre, symmetric, nuclear molecule is $^{33}$S. $^{33}$S can be considered as a 2:1 deformed nucleus with two $^{16}$O clusters with a delocalised neutron. In the DHO model, where the oscillation frequency in the $z$-direction is half that in the $x$ or $y$ directions, the orbitals occupied linked to $^{32}$S are $[n_x,n_y,n_z]=$
[0,0,0], [0,0,1], [0,0,2], [0,1,0], [1,0,0], [0,0,3], [0,1,1] and [1,0,1].
The orbitals available, associated with the next shell closure, for the valence neutron are 
[0,0,4], [0,1,2], [1,0,2], [0,2,0], [2,0,0] and [1,1,0], these are all degenerate. 
In the two-centre description, each $^{16}$O cluster occupies the orbitals
[0,0,0], [1,0,0], [0,1,0] and [0,0,1],
which leaves the following orbitals available for the valence neutron centred on each cluster $[n_x,n_y,n_z]=$ 
[0,0,2], [0,2,0], [2,0,0], [0,1,1], [1,0,1] and [1,1,0].

Given that the neutron could exist at either cluster centre with equal probability,  the coefficients associated with the linear combination of the orbitals are $1/\sqrt{2}$ (Eqn. 15). Using the 'rules' described earlier, 
the resulting two-centre molecular wave-functions can be matched to the DHO orbitals with a one-to-one correspondence:
\begin{eqnarray}
 \left[0,0,4\right] & \equiv & \frac{1}{\sqrt{2}} \Big[ [0,0,2] + [0,0,2] \Big], \nonumber \\
 \left[0,1,2\right] & \equiv & \frac{1}{\sqrt{2}} \Big[ [0,1,1] + [0,1,1] \Big], \nonumber \\
 \left[1,0,2\right] & \equiv & \frac{1}{\sqrt{2}} \Big[ [1,0,1] + [1,0,1] \Big], \nonumber \\
  \left[0,2,0\right] & \equiv & \frac{1}{\sqrt{2}} \Big[ [0,2,0] + [0,2,0] \Big], \nonumber \\
  \left[2,0,0\right] & \equiv & \frac{1}{\sqrt{2}} \Big[ [2,0,0] + [2,0,0] \Big], \nonumber \\
  \left[1,1,0\right] & \equiv & \frac{1}{\sqrt{2}} \Big[ [1,1,0] + [1,1,0] \Big]. \nonumber \\ \label{eq17}
\end{eqnarray}
Here the $n_z$ quantum number in the DHO is the sum of the two-centre $n_z$ quantum numbers and the $n_x$ and $n_y$ values remain unchanged. All of these linear combinations correspond to the addition (+) of the two-centre orbitals. The negative ($-$) possibility produces the DHO $n_z$ quantum number 2$n_z$+1, i.e. an additional node in the wave-function and hence higher energy orbitals, which can be matched to DHO counterparts in the next oscillator shell at this deformation; but not shown here.
\begin{figure}
  \begin{center}
  	\vspace{-4cm}
  \includegraphics[width=1.\textwidth]{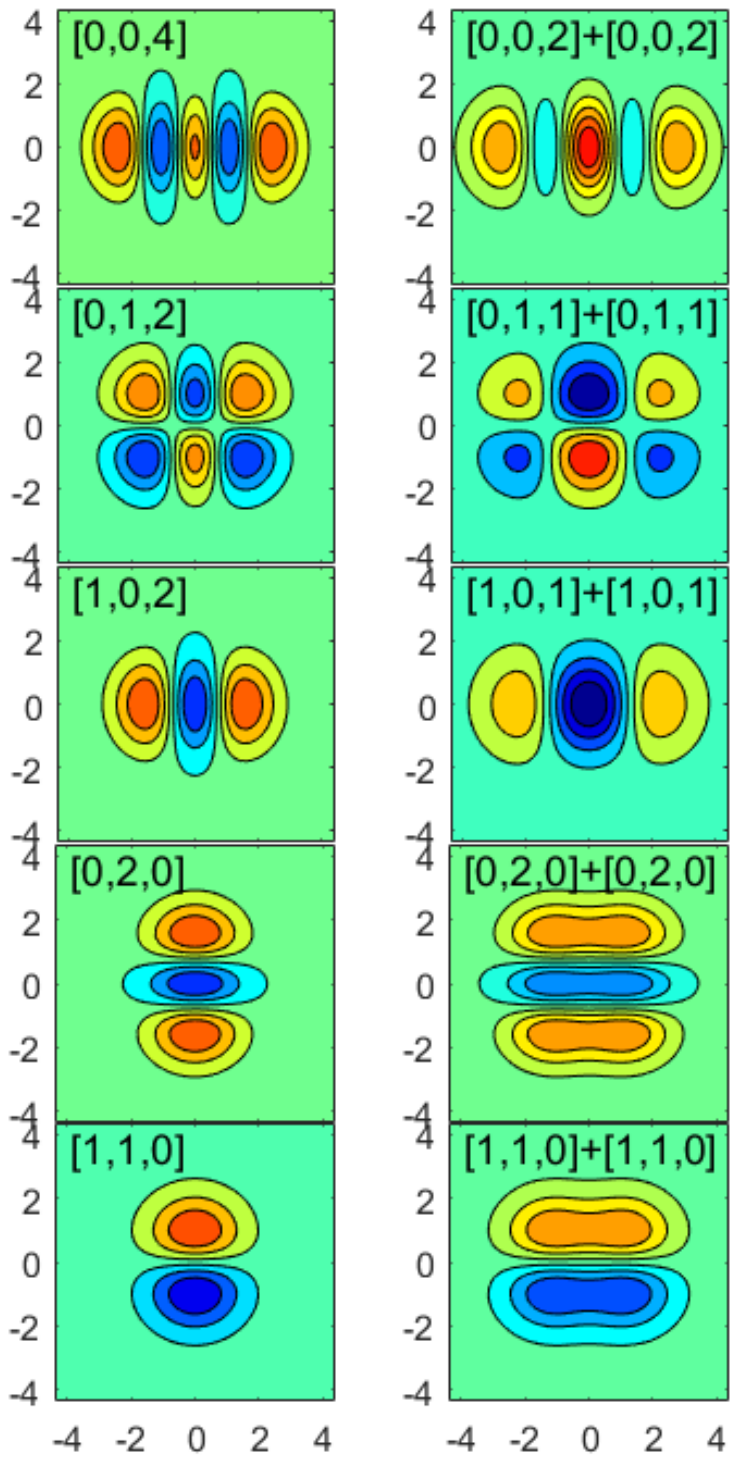}
    \vspace{-6cm}
      \caption{Comparison between 5 of the 6 possible  molecular orbitals for the valence neutron in $^{33}$S calculated in the DHO \emph{(left)} and the multi-centre, H\"{u}ckle, method \emph{(right)}. The [2,0,0] orbital has not been shown. The patterns in the wave-function amplitudes match, demonstrating that the DHO valence orbitals all have a molecular multi-centre nature. The orbitals are labelled by their harmonic oscillator quantum numbers $[n_x,n_y,n_z]$.}
    \label{fig:33S}
  \end{center}
\end{figure}
 Fig.~\ref{fig:33S}  shows the corresponding wave-functions. 
 
 In each case there is a correspondence between the patterns of the wave-function in terms of shape and nodes, though of cause there is not a precise agreement. For example, in the case of the [0,0,4] DHO wave-function, there is good agreement with the molecular wave-function calculated from the linear combination of the two axially displaced [0,0,2] orbitals. Here the final $n_z$ quantum number, 4, is the sum of the two separated $n_z$ quantum numbers, 2+2. This agreement across all molecular wave-functions demonstrates the molecular character of the valence orbitals available for the neutron in $^{32}$S above the shell gap associated with the $^{16}$O+$^{16}$O cluster structure.

The next symmetric two-centred molecule to be examined is $^{81}$Zr, which can be thought of as two $^{40}$Ca clusters with a single delocalised neutron. $^{40}$Ca is the cluster formed in the spherical HO model when orbitals are filled to the next shell closure, after that of the $^{16}$O cluster. In the 2:1 DHO model, the orbitals filled with nucleons from the $^{80}$Zr core are $[n_x,n_y,n_z]=$
[0,0,0], [0,0,1], [0,0,2], [0,1,0], [1,0,0], [0,0,3], [0,1,1], [1,0,1], [0,0,4], [0,1,2], [1,0,2], [0,2,0], [2,0,0], [1,1,0], [0,0,5], [0,1,3], [1,0,3], [0,2,1], [2,0,1] and [1,1,1]. The next highest energy orbitals available for the delocalised neutron are then
[0,0,6], [0,1,4], [1,0,4], [1,1,2], [0,2,2], [2,0,2], [0,3,0], [3,0,0], [1,2,0] and [2,1,0]. These are all degenerate at the 2:1 deformation. 
In the two-centre approach, each $^{40}$Ca cluster occupies the orbitals:
[0,0,0], [1,0,0], [0,1,0], [0,0,1], [2,0,0], [0,2,0], [0,0,2], [1,1,0], [1,0,1] and [0,1,1].
The valence neutron then occupies the orbitals
[0,0,3], [0,3,0], [3,0,0], [1,1,2], [1,0,2], [1,1,1], [0,2,1], [2,0,1], [1,2,0] and [2,1,0] around each centre.

\begin{figure}
  \begin{center}
  	\vspace{-4cm}
  \includegraphics[width=1\textwidth]{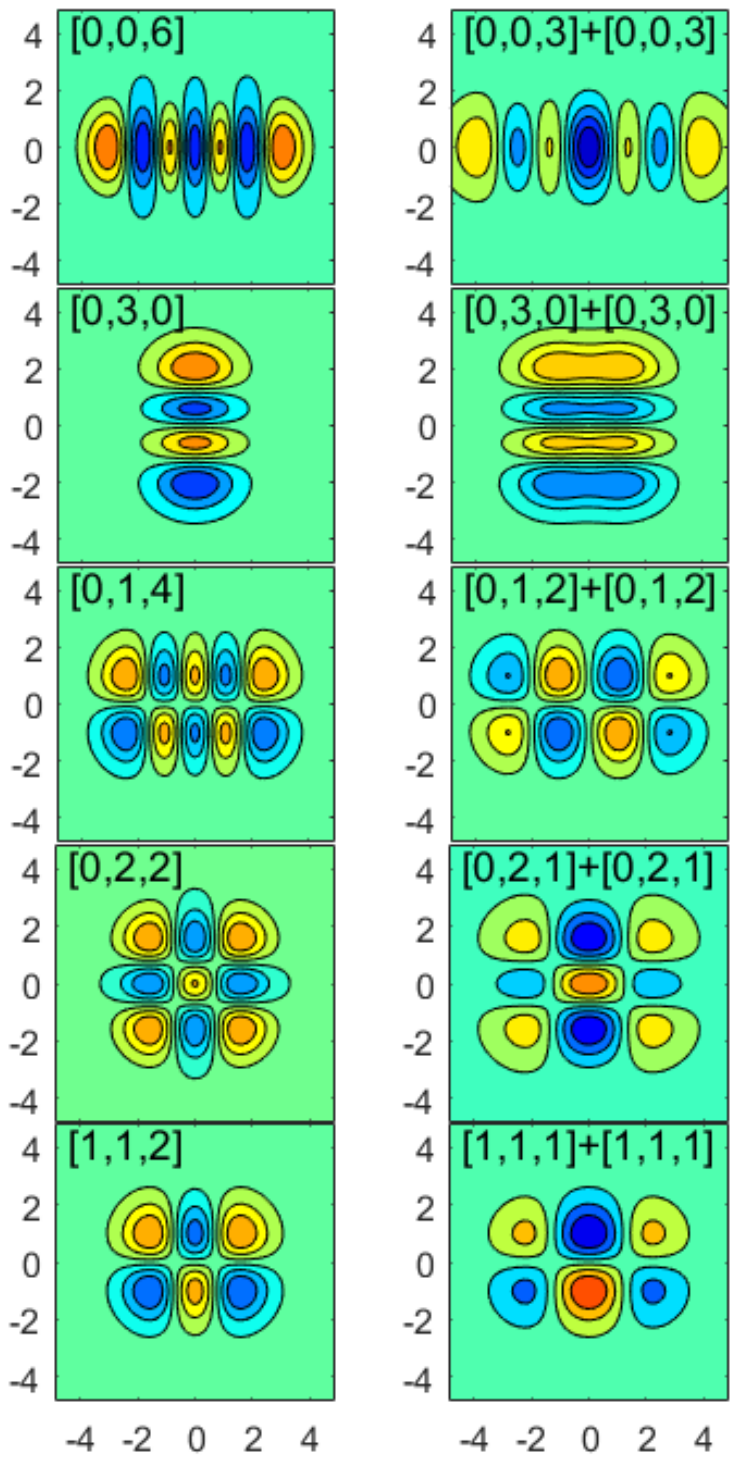}
    \vspace{-6cm}
      \caption{Comparison between selected molecular orbitals (5 of the 10 available possibilities) for the valence neutron in $^{81}$Zr calculated in the DHO \emph{(left)} and the multi-centre, H\"{u}ckle, method \emph{(right)}. The patterns in the wave-function amplitudes match, demonstrating that the DHO valence orbitals have a molecular multi-centre nature. The orbitals are labelled by their harmonic oscillator quantum numbers $[n_x,n_y,n_z]$.}
    \label{fig:81Zr}
  \end{center}
\end{figure}

As the delocalised neutron in $^{81}$Zr exists at a much higher energy than $^{33}$S, the degeneracy is much larger. This means that there are a greater number of molecular orbitals which must be constructed using linear combinations, but there is still only one possible combination of wave-functions from each cluster centre which give the correct orbital shape, only a subset of these orbitals are shown in Fig.~\ref{fig:81Zr}, but all combinations have been checked. Once again, it can be seen there is a good match between the properties of the 2:1 deformed orbitals and the explicit two-centre molecular counterparts. These two examples demonstrate that at a deformation of 2:1 the deformed harmonic oscillator orbitals above the deformed shell closure all have a molecular structure and thus the deformed harmonic oscillator explicitly includes the 2-centre molecular behaviour.

\section{Three, and More, Centre Nuclear Molecules}
In this section we will use selected examples to show that the behaviour found at 2:1 is reproduced at a deformation of 3:1 and that the inclusion of orbitals with a molecular character is more widely embedded within the deformed harmonic oscillator.

The most simple nuclear molecule with three centres is the collinear arrangement of $^{12}$C, comprising three alpha clusters. The nucleus can also be described with a 3:1 deformation in the DHO model, where the oscillation frequency along the $z$-direction is a third of that along the $x$ or $y$ directions. Following the approach in the previous section, the orbitals filled, by the core nucleons, when the nucleus is considered to exist at a single centre are [0,0,0], [0,0,1] and [0,0,2]. The next available orbitals, to be occupied by the valence neutron in $^{13}$C are
[0,0,3], [0,1,0], [1,0,0].

When the $^{13}$C nucleus is considered as three separate alpha particles, bound by a delocalised neutron, each alpha particle's nucleons occupy orbital [0,0,0]. So, the neutron could be found in a $p$-state above any of the three centres, and correspond to the degenerate 
[1,0,0], [0,1,0] and [0,0,1] spherical HO orbits.
\begin{figure}[htb]
  \begin{center}
  	\vspace{-2cm}
  \includegraphics[width=0.5\textwidth]{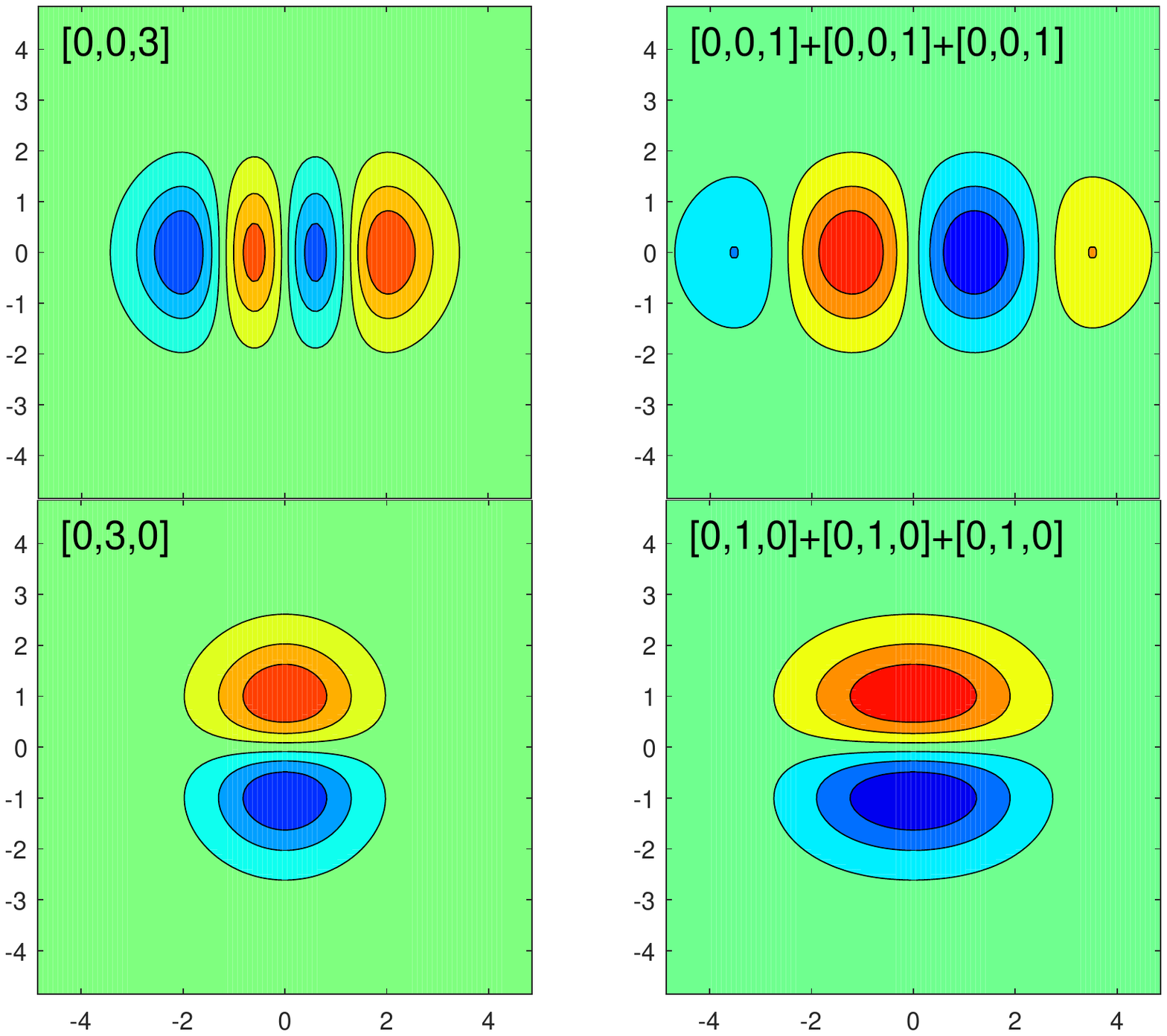}
\vspace{-3cm}
      \hspace{-0.cm}
      \caption{Amplitudes of two of the three wave-functions for the orbits occupied by valence neutrons in $^{13}$C for \emph{left}) the deformed HO and \emph{right}) the multi-centre orbits.}
    \label{fig:13C}
  \end{center}
\end{figure}
\paragraph{}The coefficients found using the H\"{u}ckel method, when linear combinations of these neutron wave-functions were made, the overall molecular wave-functions possess the form:
\begin{equation}
\psi_1  =  \frac{1}{2}\  \phi_{1} + \frac{1}{\sqrt{2}} \ \phi_{2} + \frac{1}{2} \ \phi_{3} \\
\end{equation}
\begin{equation}
\psi_2  =  \frac{1}{\sqrt{2}} \ \phi_{1} - \frac{1}{\sqrt{2}} \ \phi_{3} \\ \label{eq:mid}
\end{equation}
\begin{equation}
\psi_3  =  \frac{1}{2}\  \phi_{1} - \frac{1}{\sqrt{2}} \ \phi_{2} + \frac{1}{2} \ \phi_{3} \\
\end{equation}


The quantum numbers parallel to the deformation axis, $n_z$, are summed from each centre to give the overall number; the quantum numbers perpendicular to the deformation direction, $n_x$ and $n_y$, were conserved. The results are shown in Fig.~\ref{fig:13C}.

The lowest energy configuration for the [0,0,3] orbital comes from adding the three oscillation quanta situated at each cluster centre in the $z$-direction. Here the phases of the three wave-functions are chosen to match in the overlap region maintaining the number of total nodes in the composite wave-function as 3$\times n_z$.
There is another linear combination in which the central wave-function gives no contribution, Eqn.~\ref{eq:mid}, which produces a wave-function which resembles that of the DHO [0,0,3] orbital. Here an additional node is created in the composite wave-function by the opposite phase of the two wave-functions, generating an additional node and an additional HO oscillator quantum; $2n_z+1$.
For the [0,1,0] orbital the combination of three  oscillator quanta in the $y$-direction yields one overall oscillator quanta perpendicular to the deformation axis. This corresponds to the 3 single-centre wave-functions being added such that the phases combine constructively with no additional nodes in the $z$-direction.

\begin{figure}[htb]
  \begin{center}
  	\vspace{-4cm}
  \includegraphics[width=1\textwidth]{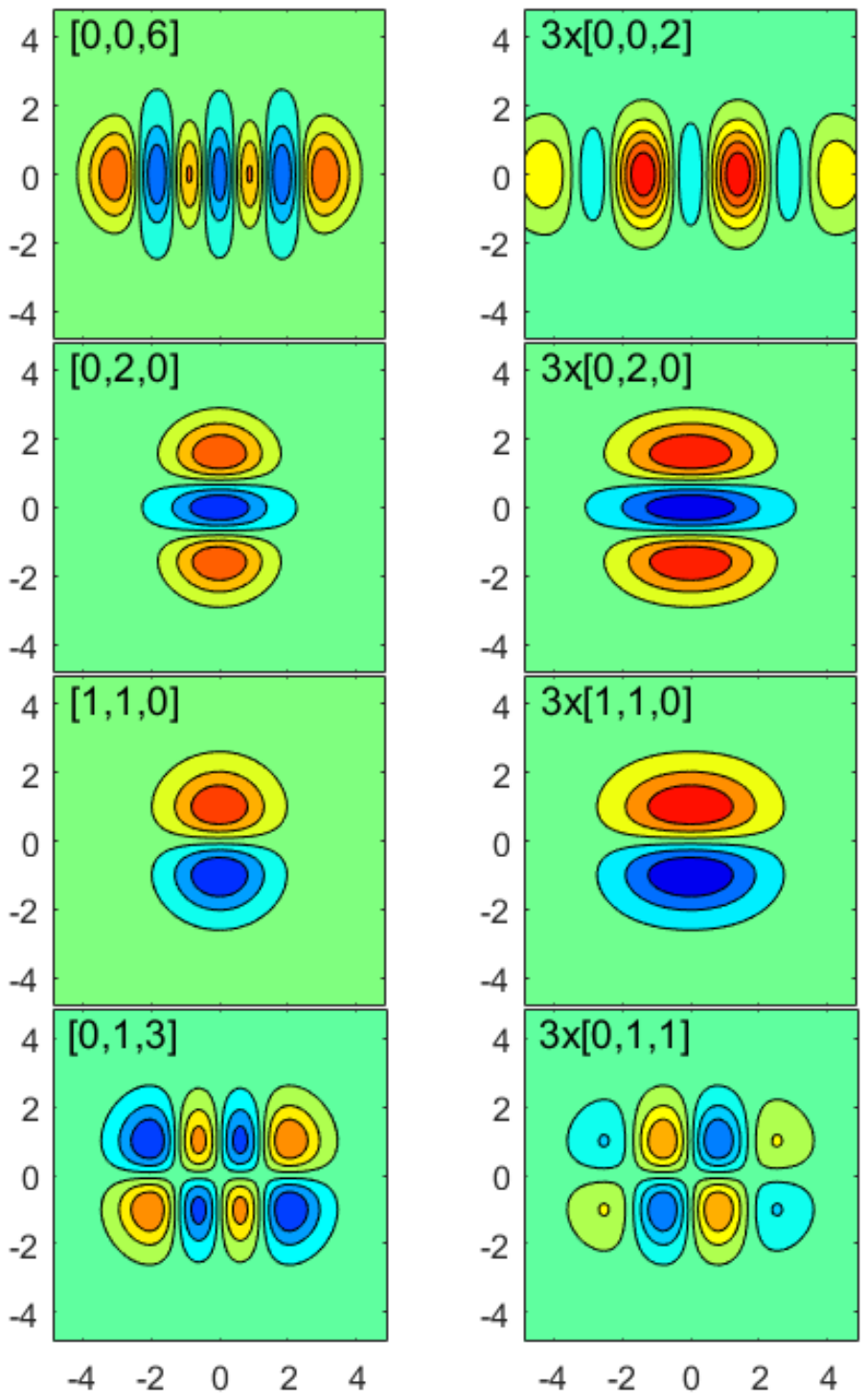}

   \vspace{-6cm}
      \hspace{-0.cm}
      \caption{Amplitudes of the wave-functions for 4 of the 6 possible  orbitals occupied by valence neutrons in $^{49}$Cr for \emph{left}) the deformed HO and \emph{right}) the multi-centre orbitals.}
    \label{fig:Cr49}
  \end{center}
\end{figure}

The next three-centre (or 3:1 deformed) nucleus with 3 $^{16}$O cores and a valence neutron is $^{49}$Cr. The possible orbitals in which the valence neutron may reside are
[0,0,6], [0,2,0], [2,0,0], [1,1,0], [0,1,3] and [1,0,3]. Fig.~\ref{fig:Cr49} shows the comparison of 4 of the 6 wave-functions constructed from the DHO and the linear combinations of the neutron wave-function at each of the $^{16}$O clusters. Again, there is a match between the properties of the valence orbitals for the covalent neutron in both the single centred, DHO, and multi-centred, H\"{u}ckle, approaches.

The structures at a deformation of 4:1 have also been explored in exactly the same way as at 2:1 and 3:1 and the same conclusions emerge; namely there is a 1 to 1 match between the nature of the wave-functions associated with the next available energy levels beyond closing the cluster cores and those predicted in the multi-centred approach. In other words, for \emph{all} structures with identical cluster cores the DHO contains the molecular symmetries for the valence neutrons.

\section{Asymmetric Nuclear Molecules}
Up to this stage, the nuclear molecules considered all have clusters of the same type, e.g. $^{32}$S=$^{16}$O+$^{16}$O and $^{80}$Zr=$^{40}$Ca+$^{40}$Ca clusters. Within the DHO energy level diagram (Fig.~\ref{fig:HO}), this means that the nuclear core will fill orbitals at each centre with the same degeneracies, e.g. 2, 6, 12, ... , and same energies; the valence neutron must then exist in a shell closure which has a new value of degeneracy. 
However, nuclear molecules which have a different types of cluster may appear in many systems, e.g. $^{20}$Ne$\equiv^{16}$O+$\alpha$. As pointed out in~\cite{voe06}, it is also possible to construct molecular orbitals from the valence neutron orbital around the two cores, e.g. from $^5$He and $^{17}$O. In this instance, the simplest nuclear molecule of this kind is $^{21}$Ne. 

For the nucleus $^{21}$Ne at a deformation of 2:1, the orbitals that are available to the valence neutron are
[0,0,3], [0,1,1] and [1,0,1]. In the multi-centre picture, there are different possible neutron orbitals which could be combined. At one centre the nuclear cluster is an alpha particle, and at the other centre the nuclear cluster is a $^{16}$O nucleus. The different possible orbitals at each centre which should be considered are therfore
[1,0,0], [0,1,0] and [0,0,1], in combination with [2,0,0], [0,2,0], [0,0,2], [1,1,0], [1,0,1] and [0,1,1].

\begin{figure}[htb]
  \begin{center}
  	\vspace{-4cm}
  \includegraphics[width=1\textwidth]{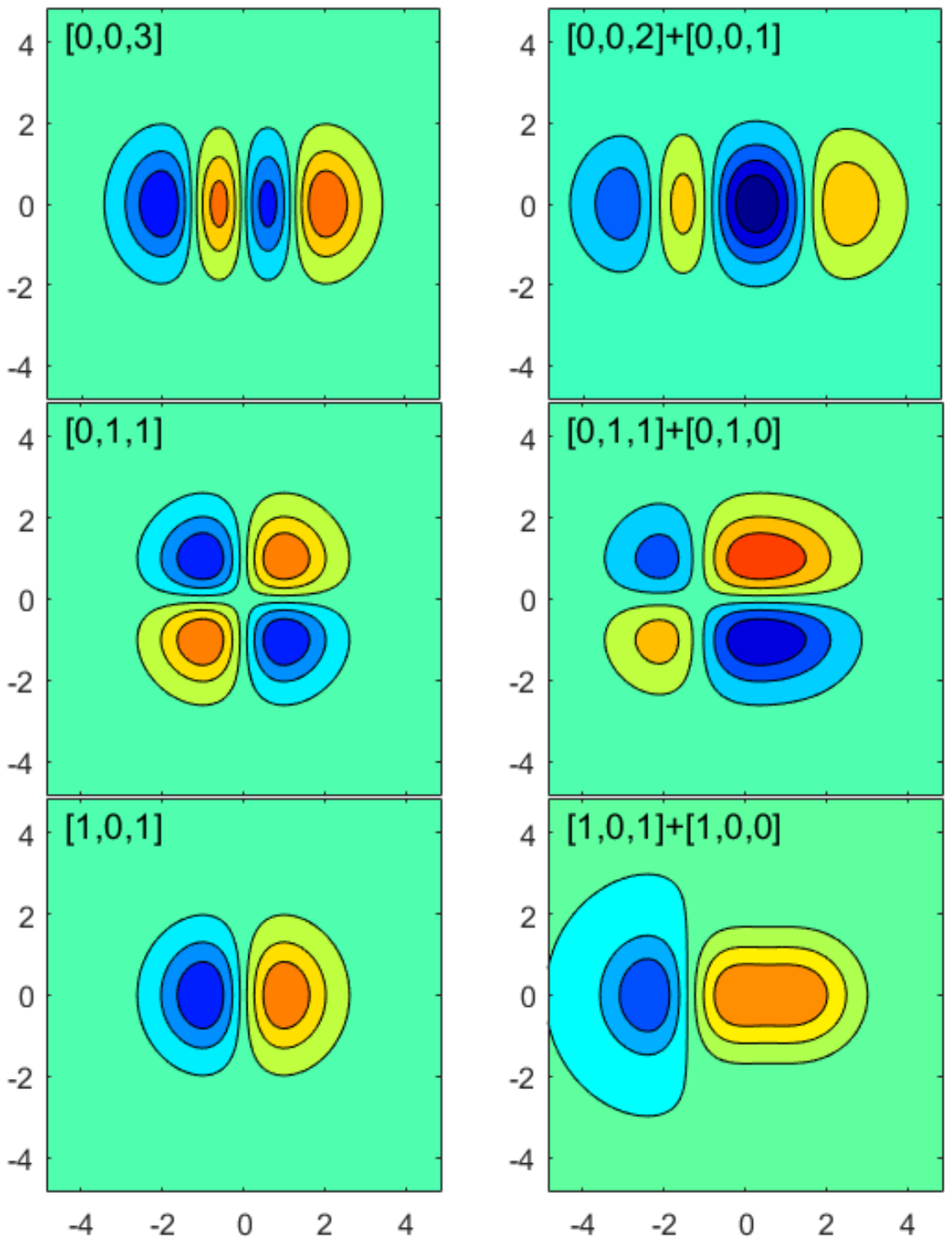}
    \vspace{-6cm}
      \hspace{-0.cm}
      \caption{Amplitudes of  orbitals occupied by valence neutrons in $^{21}$Ne for \emph{left}) the deformed HO and \emph{right}) the multi-centre orbitals, with the $^{16}$O positioned on the left hand side and the $\alpha$-particle on the right.}
    \label{fig:21Ne}
  \end{center}
\end{figure}

\begin{figure}[htb]
  \begin{center}
  	\vspace{-4cm}
  \includegraphics[width=1\textwidth]{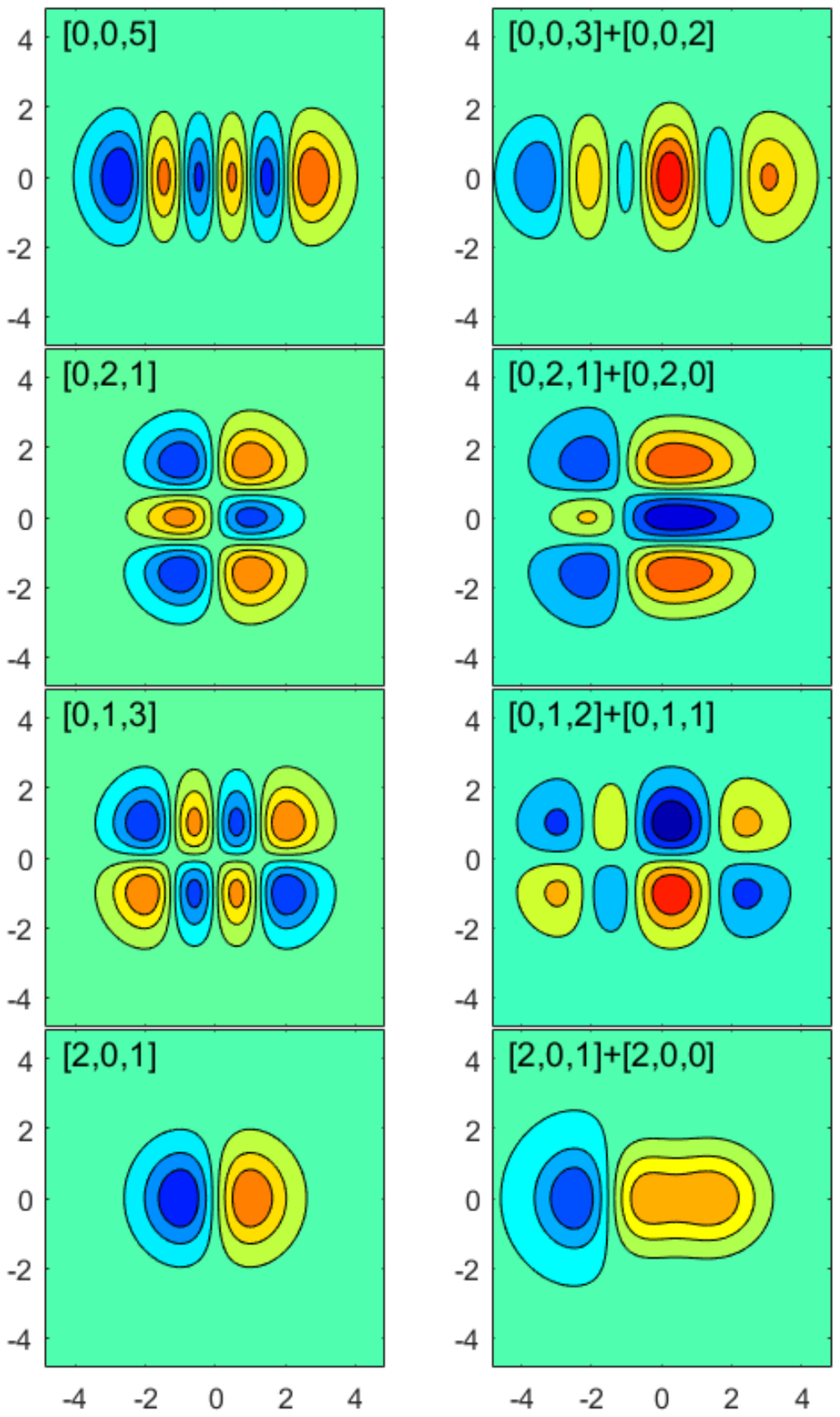}

   \vspace{-6cm}
      \hspace{-0.cm}
      \caption{Examples of the amplitudes of orbitals occupied by valence neutrons in $^{57}$Ni for \emph{left}) the deformed HO and \emph{right}) the multi-centre orbits. Four cases are shown in the top three panels. }
    \label{fig:57Ni}
  \end{center}
\end{figure}

The linear combinations made to produce the single-centre wave-function follow the same rules in terms of quantum numbers as in the case of identical cluster cores. The quantum number parallel to the direction of deformation ($z$-axis) is found by summing the $n_{z}$ values of the orbitals at each centre, and the quantum numbers perpendicular to the deformation are conserved/unchanged. However, to create wave-functions that have counterparts in the DHO, the transverse quantum numbers, in the $x$ and $y$ directions must match for the multi-centred wave-functions. So for example, as illustrated in Fig.~\ref{fig:21Ne} the linear combination of the [0,1,0] neutron orbit around the $\alpha$-cluster and [0,1,1] neutron orbit about the $^{16}$O-core results in the two centred orbit which has the representation of [0,1,1] within the DHO. If the $\alpha$-particle lies to the right-hand-side of the $^{16}$O then the [0,1,0] wave-function is superimposed on the right-hand component of the clover-like wave-function of the [0,1,1] orbit (see Fig.~\ref{fig:21Ne}).

\begin{table} 
	\caption{The relationship between the DHO orbitals for the valence neutron in the $^{40}$Ca+$^{16}$O+n, $^{57}$Ni, system and the orbitals associated with the $^{40}$Ca
		and $^{16}$O clusters.}\label{tab57Ni}
	\begin{center}
		\begin{tabular} {c|cc} \hline
Orbital in $^{57}$Ni & Orbital in $^{40}$Ca & Orbital in $^{16}$O \\
$\left[n_x, n_y, n_z\right]$ & [$n_x$, $n_y$, $n_z$] & [$n_x$, $n_y$, $n_z$] \\
\hline
$\left[0,0,5\right]$ & [0,0,3] & [0,0,2] \\
$\left[1,0,3\right]$ & [1,0,2] & [1,0,1] \\
$\left[0,1,3\right]$ & [0,1,2] & [0,1,1] \\
$\left[2,0,1\right]$ & [2,0,1] & [2,0,0] \\
$\left[0,2,1\right]$ & [0,2,1] & [0,2,0] \\
\hline
		\end{tabular}
	\end{center}
\end{table}

The $^{16}$O+$\alpha$ system is mass-asymmetric and correspondingly the resulting molecular wave-functions do not possess good parity. Parity can be restored by creating linear combinations of the molecular wave-functions formed from having the $\alpha$-particle on either the left or right side of the $^{16}$O core. The resulting wave-functions then more closely resemble the symmetric wave-functions of the deformed harmonic oscillator.    

Another asymmetric two-centre nuclear molecule that was investigated was $^{57}$Ni, which could be thought of as  $^{40}$Ca and  $^{16}$O clusters, with a delocalised neutron, at a deformation of 2:1 in the deformed harmonic oscillator. Again, it is possible to construct DHO orbitals from combinations of the single-centre orbitals around each cluster. For example, the valence [0,0,5] DHO orbital in $^{57}$Ni at 2:1 deformation is composed of the [0,0,3] and [0,0,2] orbitals around the $^{40}$Ca and $^{16}$O cores, see Fig.~\ref{fig:57Ni}.  The mapping of the molecular orbitals in $^{57}$Ni to those around the $^{40}$Ca and $^{16}$O clusters in shown in Table \ref{tab57Ni}.

It is also possible to build nuclear molecules from unequal clusters but into symmetric systems. For example, $^{25}$Mg can be constructed from the clusters $\alpha$+$^{16}$O+$\alpha$ with a valence neutron. Such systems were also explored and found that again the wave-functions formed from the linear combination of those at the three centres created the available valence orbital for the neutron in the single centre. Indeed, the ordering of the clusters did not need to be symmetric to achieve this result. Examples explored include  $^{73}$Kr (two $^{16}$O clusters and a $^{40}$Ca cluster) and $^{97}$Cd (two $^{40}$Ca clusters and a $^{16}$O cluster).
\begin{table} 
	\caption{The relationship between the DHO orbitals for the valence neutron in the $^{16}$O+$^{40}$Ca+$^{16}$O+n, $^{73}$Kr, system and the orbitals associated with the $^{40}$Ca
		and $^{16}$O clusters.} \label{tab73Kr}
	\begin{center}
		\begin{tabular} {c|ccc} \hline
			Orbital in $^{73}$Kr & Orbital in $^{16}$O & Orbital in $^{40}$Ca & Orbital in $^{16}$O \\
			$\left[n_x, n_y, n_z\right]$ & [$n_x$, $n_y$, $n_z$] & [$n_x$, $n_y$, $n_z$]& [$n_x$, $n_y$, $n_z$] \\
			\hline
			$\left[0,0,7\right]$& [0,0,2] & [0,0,3] & [0,0,2] \\
			$\left[1,0,4\right]$& [1,0,1] & [1,0,2] & [1,0,1] \\
			$\left[0,1,4\right]$& [0,1,1] & [0,1,2] & [0,1,1] \\
			$\left[2,0,1\right]$& [2,0,0] & [2,0,1] & [2,0,0] \\
			$\left[0,2,1\right]$& [0,2,0] & [0,2,1] & [0,2,0] \\
			\hline
		\end{tabular}
	\end{center}
\end{table}
The orbitals for the $^{73}$Kr case are mapped in Table~\ref{tab73Kr}. The result is very similar to that in the case of $^{57}$Ni. 

\section{Discussion and Conclusion}

This extensive study of the mapping of molecular orbitals for symmetric and asymmetric cluster systems has found that in every case the molecular orbitals have a one-to-one correspondence with the valence orbitals for neutrons at the deformation given by the underlying cluster structure. In other words, molecular states are not just found for a few select systems but are a ubiquitous feature of nuclear systems where there is an underlying cluster symmetry. 

The connection between the deformed harmonic oscillator and the appearance of molecular structures can be traced to the symmetries of the deformed harmonic oscillator~\cite{Naz92} and the ideas of Harvey~\cite{Har75}. The spherical harmonic oscillator possesses a SU(3) symmetry and it has been demonstrated ~\cite{Naz92} that at a deformation of 2:1 and 3:1, etc..., that the deformed harmonic oscillator can be described by 2 or 3 coupled SU(3) groups and hence at a deformation of N:1 the appropriate symmetry is given by N$\otimes$SU(3). In other words, the cluster structures are explicitly encoded in the symmetries. This is seen in the degeneracy patterns in Fig.~\ref{fig:HO} and the densities calculated in Figs.~\ref{fig:dho1} to~\ref{fig:dho4} and Ref.~\cite{Fre95}. 
\begin{figure}
	\begin{center}
		\vspace{-0cm}
		\includegraphics[width=0.7\textwidth]{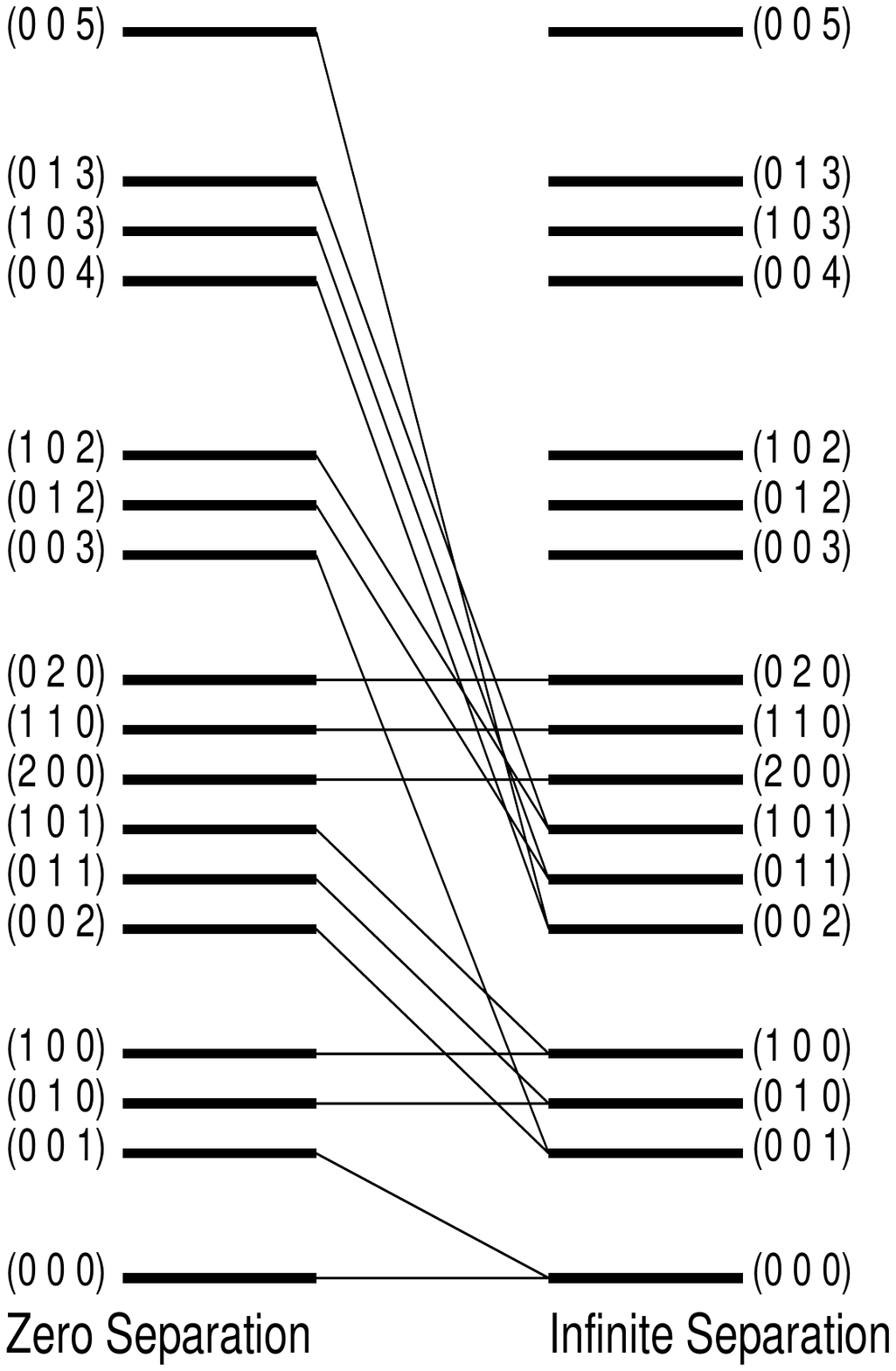}
		
		\vspace{-2cm}
		\hspace{-0.cm}
		\caption{The Harvey-diagram which demonstrates the connection between the orbitals of the single centre and two-centre system. Infinite separation is when the two clusters are far apart and zero separation is when they are fused.}
		\label{fig:Harv}
	\end{center}
\end{figure}

Harvey~\cite{Har75} noted the rules for combining two oscillator potentials which have been widely used to understand the nature and decay of cluster structures~\cite{Fre07}. These rules relate to the combination of wave-functions from the two centres to the single centre and recognised it is possible to either preserve the total number of nodes in the wave-function or create an additional node at the point where the two wave-functions connect. This means for two wave-functions with harmonic oscillator quantum numbers $n_z$ then two new wave-functions are produced with quantum numbers $2n_z$ and 2$n_z+1$. These rules are depicted in the Harvey Diagram in Fig~\ref{fig:Harv}. For three centres, merging three wave-functions with the same $n_z$ values, produces three possible values of $3n_z$, 3$n_z+1$ and 3$n_z+2$. This can be further extended to N-centres.

For symmetric clusters, the mapping of the clusters observed in the deformed harmonic oscillator to their multi-centre counterparts fully accounts for all of the levels in the separated components and the levels up to the associated shell closure in the deformed harmonic oscillator. The valence orbits above the deformed shell closure map to those outside the closed spherical core.

The merging of the two centres into a single-centre oscillator is a process which respects the Pauli Exclusion Principle, PEP, inasmuch that one of the two merging levels with identical quantum number has an additional node generated. This is also the case for the merging of multi-centres. This effect results in the molecular orbitals not being the low-lying orbitals, which tend to be occupied by the core nucleons, and hence are at higher single-particle energy. The fact that the molecular, and indeed cluster, structures are consistent with the PEP is confirmed through the AMD calculations which explicitly respect the PEP. 

For two identical clusters, e.g. two $^{16}$O nuclei, the set of degenerate levels outside the core (e.g. 6 levels in $^{16}$O) map two sets of levels at a deformation of 2:1 split in energy by 3/5$\hbar\omega$ (see Eqn. 3). These set of levels are identical to each other aside from the addition of one unit of $n_z$ (2$n_z$ and 2$n_z$+1). These are the two degeneracies of 12 observed at a deformation of 2:1 in Fig.~\ref{fig:HO}, and are the original two-centre levels which have undergone a 2$n_z$ and 2$n_z$+1 transformation.  

Similarly, the set of levels associated with the deformed harmonic cluster structure, up to and including the shell closure, at the deformation of 2:1 are the levels of the separated $^{16}$O clusters that have also undergone the 2$n_z$ and 2$n_z$+1 transformation. So for the original $^{16}$O harmonic orbitals [0,0,0], [0,0,1], [0,1,0] and [1,0,0] these become [0,0,0], [0,0,2], [0,1,0] and [1,0,0] and [0,0,1], [0,0,3], [0,1,1] and [1,0,1]. These are again identical aside from a difference of $n_z$=1. Thus, the 2$n_z$ and 2$n_z$+1 transformation applies equally for the core and valence nucleons. This process of merging from multi-centre to single centre preserves the number of nodes in the final merged wave-function plus creates a new set of wave-functions with an additional node along the separation axis. 
These principles apply to more than two centres, e.g. 3 and 4-centre cluster structures, but the transformation will be 3$n_z$, 3$n_z$+1 and 3$n_z$+2. 

For asymmetric cores the principle is the same. The merging of the two different valence orbitals around the two different cores preserves the number of total nodes in the final wave-functions and creates a second wave-function with one additional node. These resulting orbitals are the next available orbitals in the composite system, which then have a molecular structure. Linear combinations of orbitals at the two centres should have the same number of transverse nodes to create the molecular wave-functions.

The conclusion from this work is, beyond the observation that all the shell structures in the prolate deformed harmonic oscillator can be associated with cluster structures, that the next available orbitals beyond the shell-gap have molecular structure which is associated with the linear combination of the valence nucleon orbitals around the separated cluster cores. In this way molecular structures should be a wide spread phenomenon and that the molecular behaviour in nuclei is a ubiquitous phenomenon. 

The spin-orbit interaction is known to play an important role in the breakdown of clustering, and though it is included in the effective interaction of the AMD approach~\cite{Fre18} where the molecular structures of light nuclei are reproduced, there is no guarantee that the molecular structure persists to much heavier systems described in the present work. The confirmation of the robustness of the states predicted here needs to be tested with more microscopic approaches such as those described in Ref.~\cite{Fre18}.  Moreover, there are experimental challenges for the observation of such states, given that they exist at high excitation energy and will be embedded in a sea of states of other character. This will fragment the molecular states and experimental methods such as that described in Ref.~\cite{Bai19} may be required.

This work is presently being extended to oblate deformations.

\end{document}